\documentclass[12pt,final]{elsart}

\nonstopmode

\usepackage{epsfig}
\usepackage{times}

\newcommand{\eksbib}[1] {\vspace{-1.5ex}\bibitem{#1}}
\newcommand{\ekssection}[1] {\vspace{-2ex}\section{#1}\vspace{-4ex}}

\newcommand{\figadjust} {\vspace{2ex}}

\begin{document}

\begin{frontmatter}
  
  \title{\vspace{2cm}Modifications of the hydrogen bond network \\
    of liquid water in a cylindrical {\boldmath${\rm SiO}_2$} pore}

  \author{Christoph Hartnig, Wolfgang Witschel, }
  \author{Eckhard Spohr\thanksref{comm}}
  \address{Department of Theoretical Chemistry, University of Ulm,\\
    \mbox{Albert-Einstein-Allee 11}, D-89069 Ulm, Germany}
  \author{Paola Gallo, Maria Antonietta Ricci, Mauro Rovere}
  \address{Dipartimento di Fisica, Universit{\`a} di Roma Tre, INFM,\\
  Unit{\`a} di Ricerca Roma Tre, Via della Vasca Navale 84, I-00146
  Roma, Italy }
\thanks[comm]{Author to whom correspondence should be addressed.\\
  \hspace*{6pt} Electronic mail: eckhard.spohr@chemie.uni-ulm.de}

\begin{abstract}
  We present results of molecular dynamics simulations of water
  confined in a silica pore. A cylindrical cavity is created inside a
  vitreous silica cell with geometry and size similar to the pores of
  real Vycor glass. The simulations are performed at different
  hydration levels. At all hydration levels water adsorbs strongly on
  the Vycor surface; a double layer structure is evident at higher
  hydrations.  At almost full hydration the modifications of the
  confinement-induced site-site pair distribution functions are in
  qualitative agreement with neutron diffraction experiment.  A
  decrease in the number of hydrogen bonds between water molecules is
  observed along the pore radius, due to the tendency of the molecules
  close to the substrate to form hydrogen-bonds with the hydrophilic
  pore surface. As a consequence we observe a substrate induced
  distortion of the H-bond tetrahedral network of water molecules in
  the regions close to the surface.
\end{abstract}

\end{frontmatter}

\ekssection{Introduction}
The increasing interest in the study of structural and dynamical 
properties of confined water in different environments arises from the
close connection with many relevant technological and biophysical
problems~\cite{chen}.  It is well known that the properties of liquid
water at ambient conditions are mainly determined  by the microscopic
local tetrahedral order~\cite{stanley}.  Understanding how the
connected random hydrogen bond network of bulk water is modified when
water is confined in small cavities inside a substrate material, is
very important, e.~g.,  for studies of stability and enzymatic activity of
proteins, oil recovery, or heterogeneous catalysis, where
water-substrate interactions play a fundamental role. The modifications
of the short range order in the liquid depend on the
nature of the water-substrate interaction, i.e. hydrophilic or
hydrophobic, as well as on its spatial range and on the geometry of  the
substrate. 
Many experimental studies of structural and dynamical 
properties of water in confined regimes have been 
performed~\cite{chen,mcbf1,mcbf2,mar1,mar2,jcd,nmr}.
Water-in-Vycor is one of the most extensively studied systems. 
In particular, it has been found in a recent neutron
scattering experiment~\cite{mar1,mar2} 
that the hydrogen bond network of water in Vycor
is strongly distorted with respect to bulk water.
In order to shed light on this behavior,
which was not evident in previous
experiments~\cite{mcbf2}, a more extensive
study of the 
microscopic details of the molecular arrangement is needed.
 
Computer simulation methods allow a more detailed exploration of the
molecular arrangements inside the system than most experimental
techniques. They can therefore
be considered as an ideal microscopic tool to study
the properties of water in confined geometries, subject, of course, to
the usual limitations of the method and the interaction potential
functions employed.  In this paper we present results obtained by a
computer simulation study of water confined in a model for a Vycor
glass pore.  The simulation cell is obtained by carving a single
cylindrical cavity in a modeled silica glass, thus reproducing the
symmetry, the average dimensions, and the structure of the rough
surface of Vycor pores at the atomic level.

In the next section we will briefly review the method for building the
simulation cell and the implementation of the molecular dynamics.  In
the following sections we summarize some of the key results and
compare them with recent experiments. The discussion is more detailed
than in earlier work by our groups~\cite{old,vycor1}.

\ekssection{Model for the Vycor pore and molecular dynamics calculations}

A glass of ${\rm SiO_2}$ is obtained by computer simulation, following
the method proposed by Brodka and Zerda~\cite{brodka}, using potential
parameters given by Feuston and Garofalini \cite{feuston}.  The system
is simulated as a simple ionic model for ${\rm Si}^{4+}$ and ${\rm
  O}^{2-}$ ions~\cite{c0372}.  Inside a cell of vitreous SiO$_2$ with
a box length of approximately 71 {\AA} a cylindrical cavity of 40
{\AA} diameter, corresponding to the average pore size of the Vycor
glass of interest, is created.  The details for generating the cavity
with a rough surface and microscopic structure representative for
pores in Vycor glass are given in previous
publications~\cite{vycor1,c0302}.  On the surface of the pore two
types of oxygen atoms can be distinguished, depending on the number of
silicon atoms to which they are connected.  {\em Nonbridging} oxygens
(NBO's) are bonded to only one silicon atom and are saturated with a
hydrogen atom.  The number density of the hydroxyl groups in our
samples is 2.5 $\rm nm^{-2}$, in good agreement with the 
experimentally determined value~\cite{c0252}. The second species of surface oxygen atoms
are {\em bridging oxygens} (BO's), which are connected to two silicon atoms.

For the water-water interaction we assume the SPC/E model~\cite{spce}.
The atoms of the substrate are allowed to interact with the water
sites by means of the Coulomb potential, where different fractional
charges are assigned to BO's ($-0.629 |e|$), NBO's ($-0.533 |e|$),
silicon atoms ($1.283 |e|$) and surface hydrogen atoms ($0.206 |e|$).
In addition both BO's and NBO's interact with the oxygen sites of
water via a Lennard-Jones potential, whose parameters are $\sigma
=2.70$ \AA{} and $\sigma =3.00$ \AA{} for BO and NBO atoms,
respectively, and $\varepsilon = 230$ K in both cases~\cite{brodka}.
During the simulation the Vycor atoms are kept fixed, and periodic
boundary conditions are applied along the $z$-axis. The shifted force
method is used with a cut-off at 9 {\AA}.  As discussed in a previous
paper~\cite{vycor1}, the use of larger cut-off or Ewald summation does
not change the nature of the obtained results.  The temperature is
controlled by means of a Berendsen thermostat with relaxation times
between 0.5 and 1 ps. This procedure simplifies running the
simulations over long times but has negligible effects on the
calculated properties.  All calculations have been performed at room
temperature for several hundred picoseconds, after an equilibration
run lasting more than 50 ps. More details can be found in
Ref.~\cite{vycor1}.

\ekssection{Structural properties of the confined water}
\begin{figure}[tb] 
  \begin{center}
    \epsfig{file=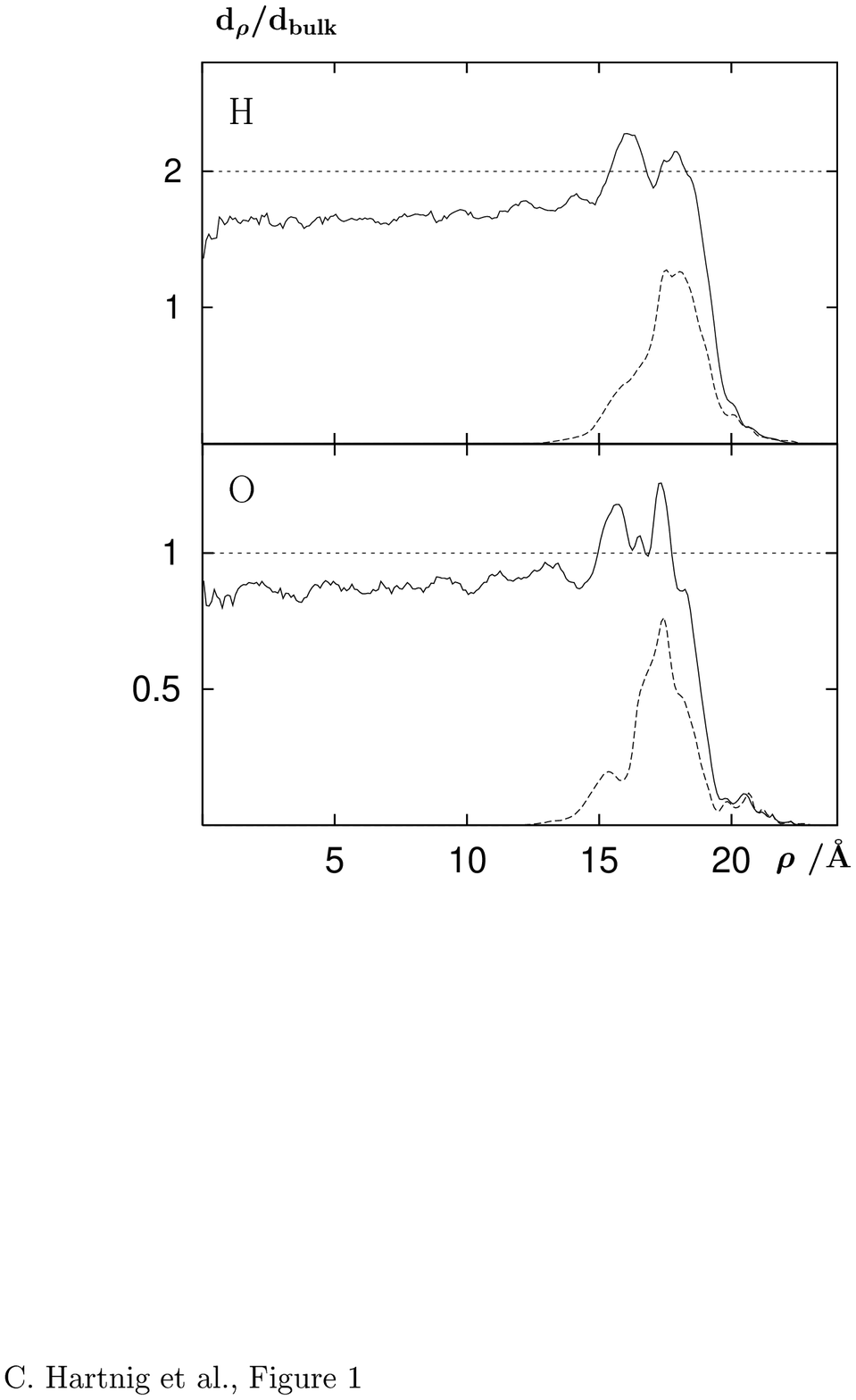,bbllx=134,bblly=382,bburx=462,bbury=780,clip=,scale=0.6}
    \caption{
      Radial density profiles (normalized to the bulk density of water at
      ambient condition) of oxygen atoms (bottom) and hydrogen atoms
      (top) from simulations at two different hydration levels:
      96 \% hydration (2600 water molecules; full line) and 
      19 \% hydration (500 water molecules; dashed line).
      }
    \label{fig:dens}
    \figadjust
    \figadjust
  \end{center}
\end{figure}
The molecular dynamics calculations have been performed for different
numbers of water molecules, corresponding to different levels of
hydration.  The density profiles of the oxygen and hydrogen atoms of
water for 96\% (2600 water) and 19\% degree of hydration (500 water)
are shown in Fig.~\ref{fig:dens}. Density profiles and further results
of several intermediate degrees of hydration have been published
elsewhere \cite{vycor1}.  We observe already at low hydration the
presence of a layer of water molecules wetting the substrate surface;
at nearly full hydration two layers of water with higher than average
density are evident.  The same features are present in the density
profiles for the hydrogen atoms.  Few molecules are able to penetrate
into the glass and are trapped inside small pockets close to the
surface, which are a byproduct of the 'sample preparation process'.
The molecules in these pockets can easily be identified in the
snapshot along the pore axis in Fig.~\ref{fig:ring}.  At low
hydrations the surface is not covered completely; adsorption occurs
preferentially in several regions, in which small clusters are formed.
We have made no attempt to identify specific surface structures that
favor cluster formation.
\begin{figure}[tb] 
  \begin{center}
    \epsfig{file=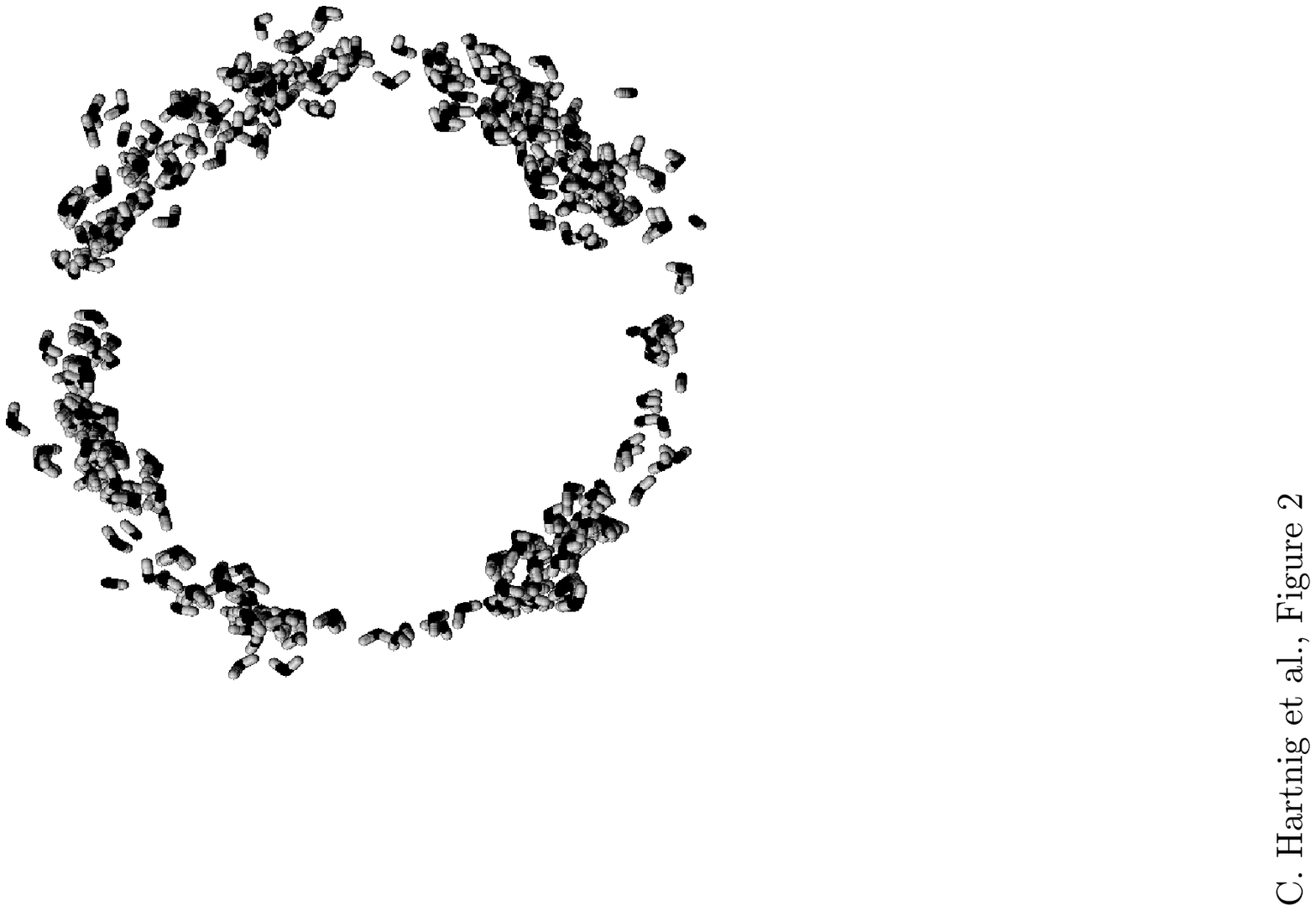,bbllx=145,bblly=369,bburx=419,bbury=652,angle=90,clip=,scale=0.7}
    \caption{Snapshot from the simulation of $500$ SPC/E water molecules.}
    \label{fig:ring}
    \figadjust
    \figadjust
  \end{center}
\end{figure}

In the comparison with experiment the definition of the hydration
level deserves some consideration. Full hydration in the experimental
sample preparation corresponds to an estimated water density which is
$11 \%$ less than the density of water at ambient
conditions~\cite{c0355}.  If the simulated cavity would be a perfect
cylinder of radius 20 \AA{}, this would translate into $N_W = 2670$
molecules.  Due to the roughness of the glass surface and the
existence of the small pockets inside the glass close to the surface,
which trap several water molecules over the entire simulation length,
the exact number of molecules can only be determined by a grand
canonical simulation scheme.  For the investigated number $N_W=2600$
we observe indeed that the density profile does not reach the bulk
value in the center of the pore. If one assumes that full hydration
corresponds to a bulk-like region in the center of the pore, we
estimate that an additional 100 molecules would correspond to full
hydration, not much different from the experimental value. Thus,
$N_W=2600$ corresponds to a hydration level of approximately 96\%.

\begin{figure}[tb] 
  \begin{center}
    \epsfig{file=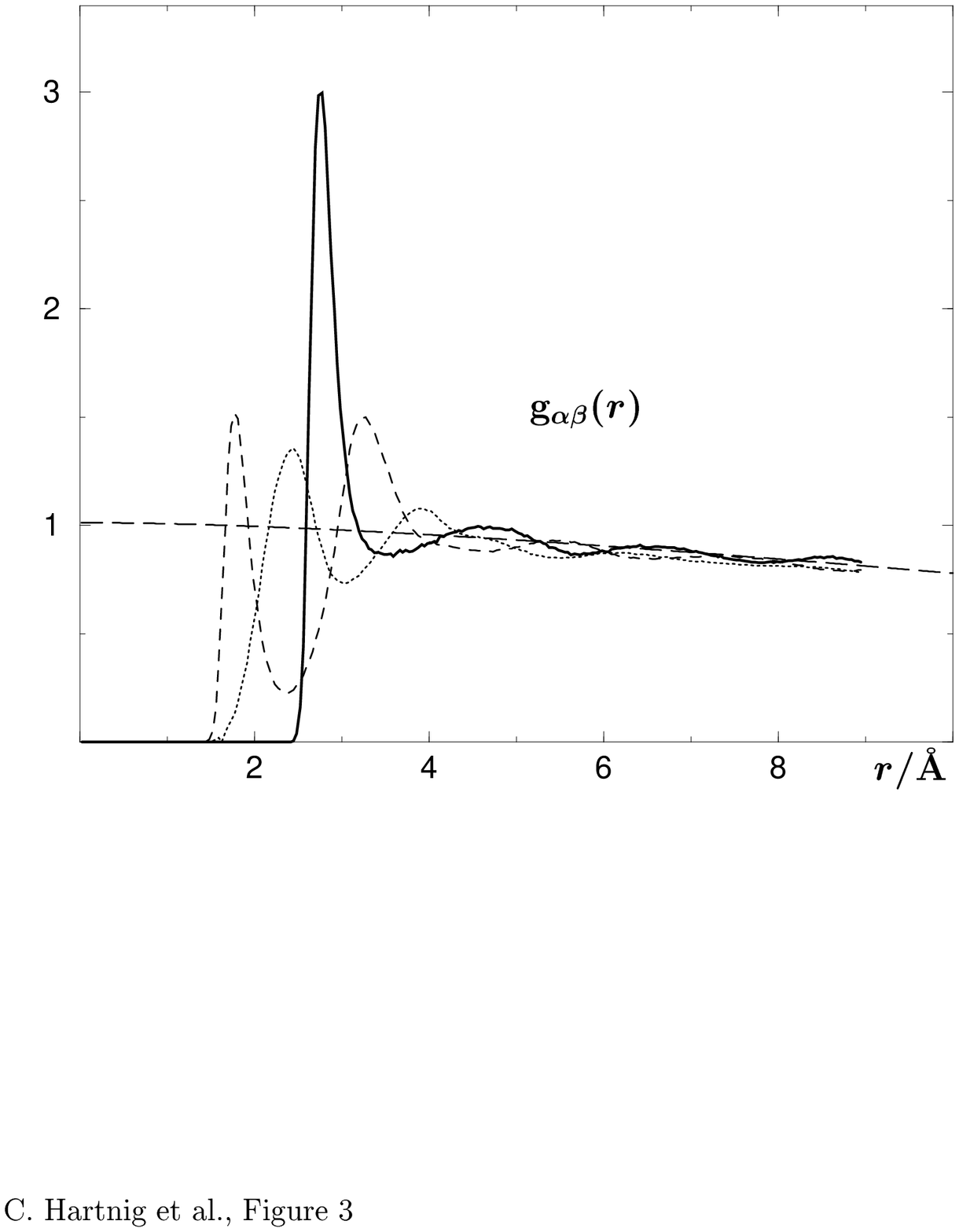,bbllx=93,bblly=350,bburx=520,bbury=720,clip=,scale=0.55}
    \caption{Raw site-site distribution function calculated without
      including finite size effects. Continous, dashed and
      short-dashed lines correspond to
      oxygen-oxygen, oxygen-hydrogen and hydrogen-hydrogen
      distribution functions, respectively. The long-dashed curve is the
      Fourier transform of the form factor of a cylinder of
      radius $20$ {\AA}.}
    \label{fig:grmd}
    \figadjust
    \figadjust
  \end{center}
\end{figure}
We calculate the site-site radial distribution functions for
$N_W=2600$, i.~e., at approximately the experimental density.  The
calculation of these functions is not straightforward when dealing
with a non-periodic system, since excluded volume effects must be
carefully taken into account~\cite{mar2,soper,branka}.  Unlike a bulk
liquid, a \emph{uniform confined fluid}, i.e., a collection of
non-interacting particles in a confining volume, where periodic
boundary conditions are only along the $z$-axis, does not create a
\emph{uniform} radial distribution function $g=1$. Rather, the form of
the radial distribution function depends on the geometry of the
confining system.

If we disregard the confinement effects, the raw site-site pair
correlation functions can be calculated in the usual way from the
simulation by taking the average number $n_{\alpha \beta}^{(2)}(r)$ of
sites of type $\beta$ lying in a spherical shell $\Delta v(r)$ at
distance $r$ from a site of type $\alpha$ and normalizing it to the
average number of atoms in the same spherical shell in an ideal gas at
the same density~\cite{allen}
\begin{equation}
g_{\alpha \beta}^{(MD)} \left( r \right) = 
\frac{ n_{\alpha \beta}^{(2)}(r) }{ {\frac{N_\beta }{V_p}} 
\Delta v(r)} \ ,
\label{grmd}
\end{equation}
where $N_\beta$ is the total number of $\beta$ sites, and $V_p$ is the
volume of the cylindrical cell. The three resulting raw site-site
correlation functions are shown in Fig.~\ref{fig:grmd}.
It is evident that the three functions
do not approach a constant value at large $r$.
The normalization factor in (\ref{grmd}) must also 
include a uniform profile,
\begin{equation}
g_u(r)=\frac{V_p}{\left( 2\pi \right)^3} \int d^3 Q P_{cyl} \left(Q \right)\,.
\label{gru}
\end{equation}
$P_{cyl}(Q)$ is the form factor of the cylindrical simulation
cell of height $L$ and radius $R$~\cite{glatter} 
\begin{equation}
P_{cyl} \left( Q \right) =
\int_0^1 d\mu \left[ j_0 \left( \frac{\mu Q L}{2} \right) \right]^2
\left[ \frac{2 j_1 \left( Q R \sqrt{1-\mu^2} \right)}{Q R\sqrt{1-\mu^2}}
\right]^2 
\label{pcyl}
\end{equation}
where $j_n(x)$ are the spherical Bessel functions of the first kind of
order $n$. The resulting $g_u(r)$ is the smooth dashed line shown in
Fig.~\ref{fig:grmd}.

\begin{figure}[tb] 
  \begin{center}
    \epsfig{file=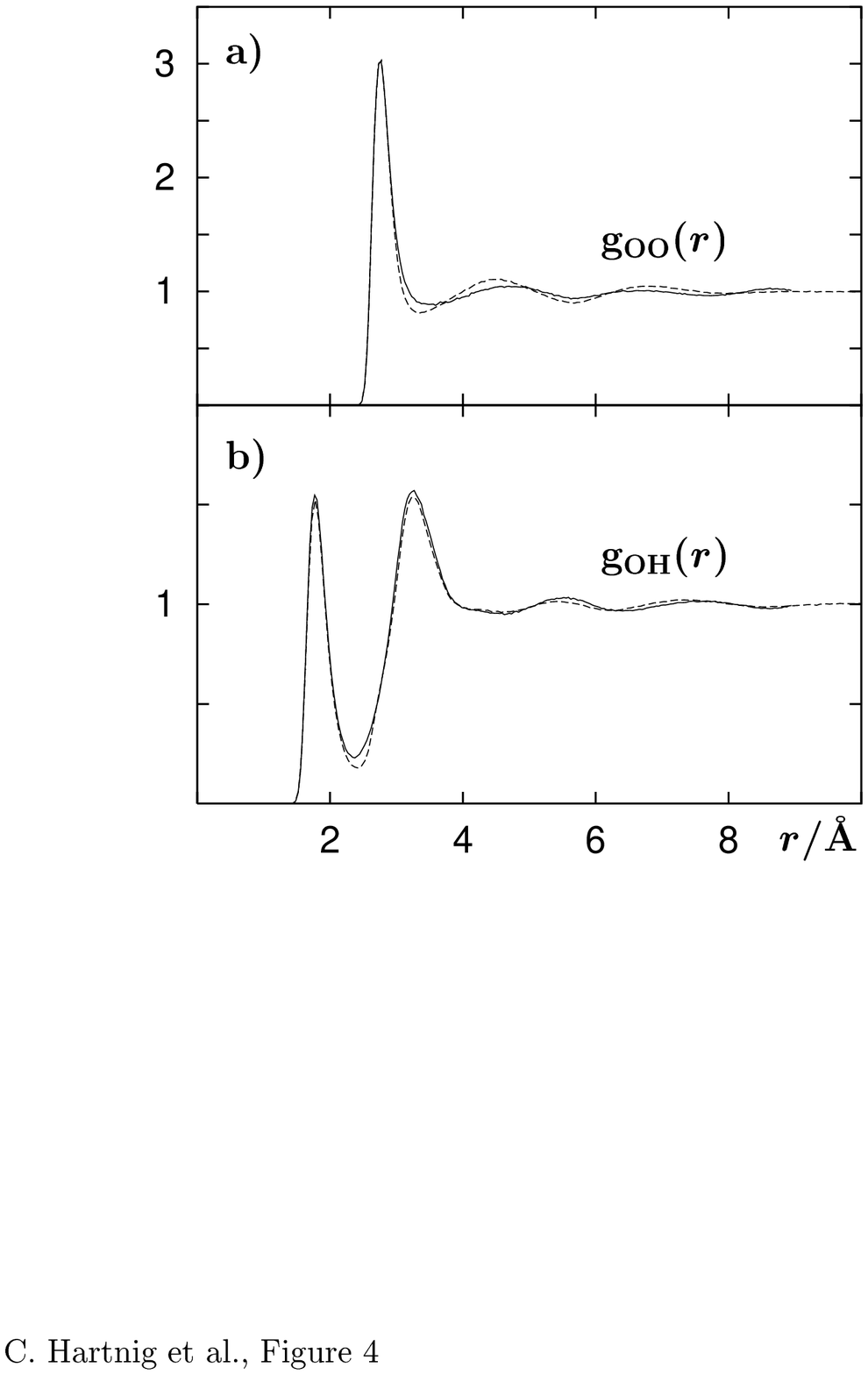,bbllx=140,bblly=368,bburx=451,bbury=742,clip=,scale=0.6}
    \caption{Computer simulated site-site distribution functions,
      for confined water (solid lines).
      These functions have been corrected taking into account excluded
      volume effects, as explained in the text, and are compared with results
      obtained for bulk SPC/E water (dashed lines). 
      a)~Oxygen-oxygen distribution function. b)~Oxygen-hydrogen
      distribution function.}
    \label{fig:gr}
    \figadjust
    \figadjust
  \end{center}
\end{figure}
The properly normalized oxygen-oxygen and oxygen-hydrogen
site-site distribution functions are then obtained by dividing
$g^{MD}_{\alpha\beta}(r)$ by $f_c \cdot g_u(r)$.
The correction factor $f_c$, which lies between 1.0 and 1.04,
takes into account the roughness of the surface and has been
adjusted in such a way that the corrected pair correlation functions
are close to one in the region $r>8$ \AA.
The corrected pair correlation functions are shown in
Fig.~\ref{fig:gr} and are compared to the corresponding functions of
SPC/E water at ambient conditions (dashed lines).  We note that the
modification of the oxygen-oxygen function relative to bulk water
shows a similar trend as the experimental one (see Fig.~7a in
Ref.~\cite{mar2}).  The first minimum becomes shallower and 
fills in.  Such behavior was also observed in experiments and computer
simulations of water under pressure, which was explained as resulting
from a collapsing hydrogen-bond network~\cite{okhulkov,a335}.  In the
oxygen-hydrogen function we notice that the first peak is lower in
amplitude and the first minimum shifts towards shorter distances, as
was observed in the experiment (Fig.~7b in Ref.~\cite{mar2}). 
Agreement with experimental results is less satisfactory in the case
of the hydrogen-hydrogen pair correlation function.  The experimental
function is changed much more than the simulated one. The differences
may be due to the fact that, in the experiment, the hydroxyl hydrogen
atoms of Vycor are not distinguished from water hydrogen atoms.
 
We conclude that our results are in qualitative agreement with the
experimental trend for the confinement-induced changes in water
structure.  According to Ref.~\cite{mar2}, those changes 
indicate a distortion of the H-bond ordering in the confined water,
induced by the interaction with the substrate.  The distortion of the
hydrogen bond network deserves further investigation. Since there are
several approximations made in extracting the water-water contribution
from the measured structure factors, computer simulations can
contribute to our understanding by 
first providing a more detailed analysis of the simulated hydrogen 
bond network than is possible in experiment, and second 
by investigating the differences between water molecules located at various
distances from the substrate surface.

\ekssection{Hydrogen bond network of confined water}
\begin{figure}[tb] 
  \begin{center}
    \epsfig{file=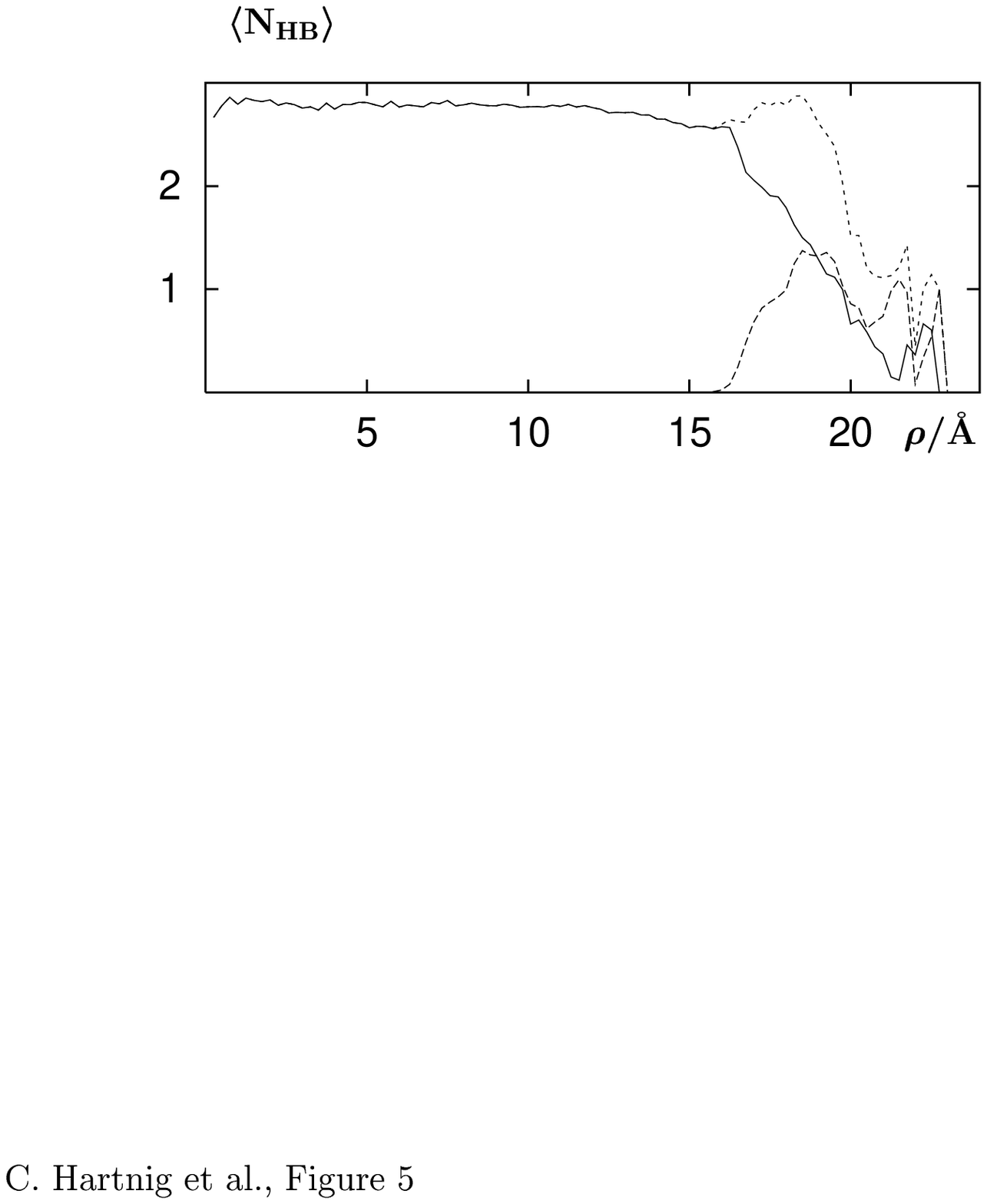,bbllx=120,bblly=524,bburx=451,bbury=707,clip=,scale=0.7}
    \caption{Number of hydrogen bonds along the
      pore radius in the system with $N_W=2600$. 
      The solid line refers to water-water hydrogen bonds, the long-dashed line corresponds
      to hydrogen bonds between water molecules and Vycor atoms and the short-dashed
      lines represents the total number of hydrogen bonds.}
    \label{fig:hb}
    \figadjust
    \figadjust
  \end{center}
\end{figure}
In a first attempt to characterize the hydrogen bond network, we
calculated the number of hydrogen bonds between water molecules and
between water molecules and surface atoms. A geometric criterion is
applied.  A hydrogen bond exists between two oxygen atoms, if the
angle between the intramolecular O-H vector and the intermolecular
O$\cdots$O vector is less than $30^\circ$, provided that the O$\cdots$O
separation is less than $3.35$ {\AA}.  In Fig.~\ref{fig:hb} we show the
results for $N_W=2600$.  Close to the surface the number of
water-water hydrogen bonds is reduced; a large number of interfacial
water molecules is engaged in hydrogen bonds between water molecules
and oxygen atoms of the pore surface.  Taking the sum of both
contributions, one can see that the total number of hydrogen bonds
remains constant and the loss of water-water hydrogen bonds is
compensated by water-Vycor bonds.  Only bridging oxygen atoms of the
Vycor surface contribute significantly to the number of hydrogen
bonds; the amount of bonds between hydroxyl groups on the surface and
water molecules is small and therefore not shown in
Fig.~\ref{fig:hb}.  It is worth noticing that at this high level of
hydration the average number of H-bonds is always lower than the
number of bonds formed by bulk SPC/E water.

\begin{figure}[tb] 
  \begin{center}
    \epsfig{file=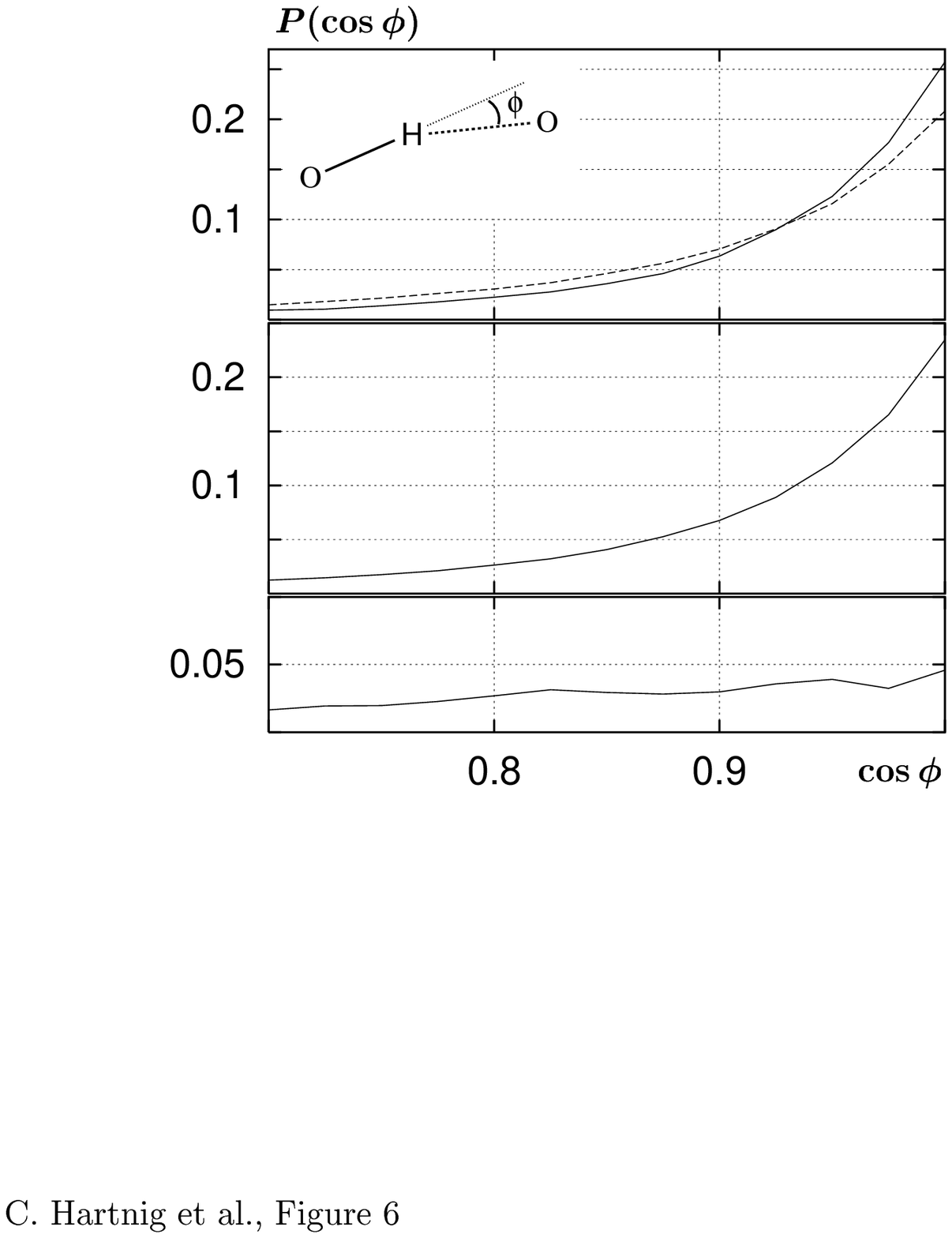,bbllx=128,bblly=411,bburx=451,bbury=736,clip=,scale=0.7}
    \caption{Hydrogen bond angle distribution function, $P(\cos \phi)$, (see inset).
      Top: nearest neighbor water pairs ($r_{\rm OO}<3.35$ \AA{})
      in the range $\rho=$5--10 {\AA} (solid line) $\rho=$15-20 {\AA} (dashed line).
      Center: the corresponding distribution between water and bridging oxygen
      atoms. Bottom: the corresponding distribution between water and non-bridging oxygen
      atoms.}
    \label{fig:phi}
    \figadjust
    \figadjust
  \end{center}
\end{figure}
In Fig.~\ref{fig:phi} we have investigated the distribution of
hydrogen bond angles in more detail for the simulation with $N_W =
2600$.  The figure depicts the distribution of $\cos\phi$, where
$\phi$ is the angle between intramolecular O-H and intermolecular H$\cdots$O
vector (see inset). Only that part of the distribution which
corresponds to almost linear hydrogen bonds is shown.  The top frame
shows the distribution of water-water hydrogen bond angles close to
the center of the cylinder (solid line) and close to the pore surface
(dashed). While there is a reduction in the number of water-water
bonds, their nature is not changed. By and large, linear water-water
hydrogen bonds are favored. There is however, a tendency to have
slightly more distorted hydrogen bonds close to the pore surface, as can be
inferred from the reduced height of the distribution 
(dashed line) at $\cos \phi =1$ and the increase for $\cos\phi < 0.9$.

The center frame of Fig.~\ref{fig:phi} contains the corresponding
distribution for the hydrogen bonds between water molecules and the
bridging oxygen atoms.  The distribution is almost identical to that
between water molecules. Hence, the loss in water-water bonds is
quantitatively (see Fig.~\ref{fig:hb}) and structurally compensated by
hydrogen bonds to Vycor atoms.  The linearity of hydrogen bonds
between water molecules and surface oxygen atoms on hydrophilic
substrates was already observed by Lee and Rossky \cite{c0263}.  The
angle distribution between water molecules and hydroxyl groups on the
surface (bottom frame of Fig.~\ref{fig:phi}) is uniform; there is no
tendency to form linear bonds between water and NBO's as in the case
of water and BO's.  The probable cause is the fact that, contrary to
Lee and Rossky's work, all substrate atoms including the hydrogen
atoms are immobile during the simulation and that, consequently, the
hydroxyl groups are not able to reorientate in an optimal way to form
hydrogen bonds with water molecules. Almost linear hydrogen bonds can,
however, form between water molecules and BO's due to the rotational
mobility of the water molecule.  Since the total number of hydrogen
bonds between water and NBO's is very small, we did not attempt to
modify our model of the Vycor surface.

\begin{figure}[tb] 
  \begin{center}
    \epsfig{file=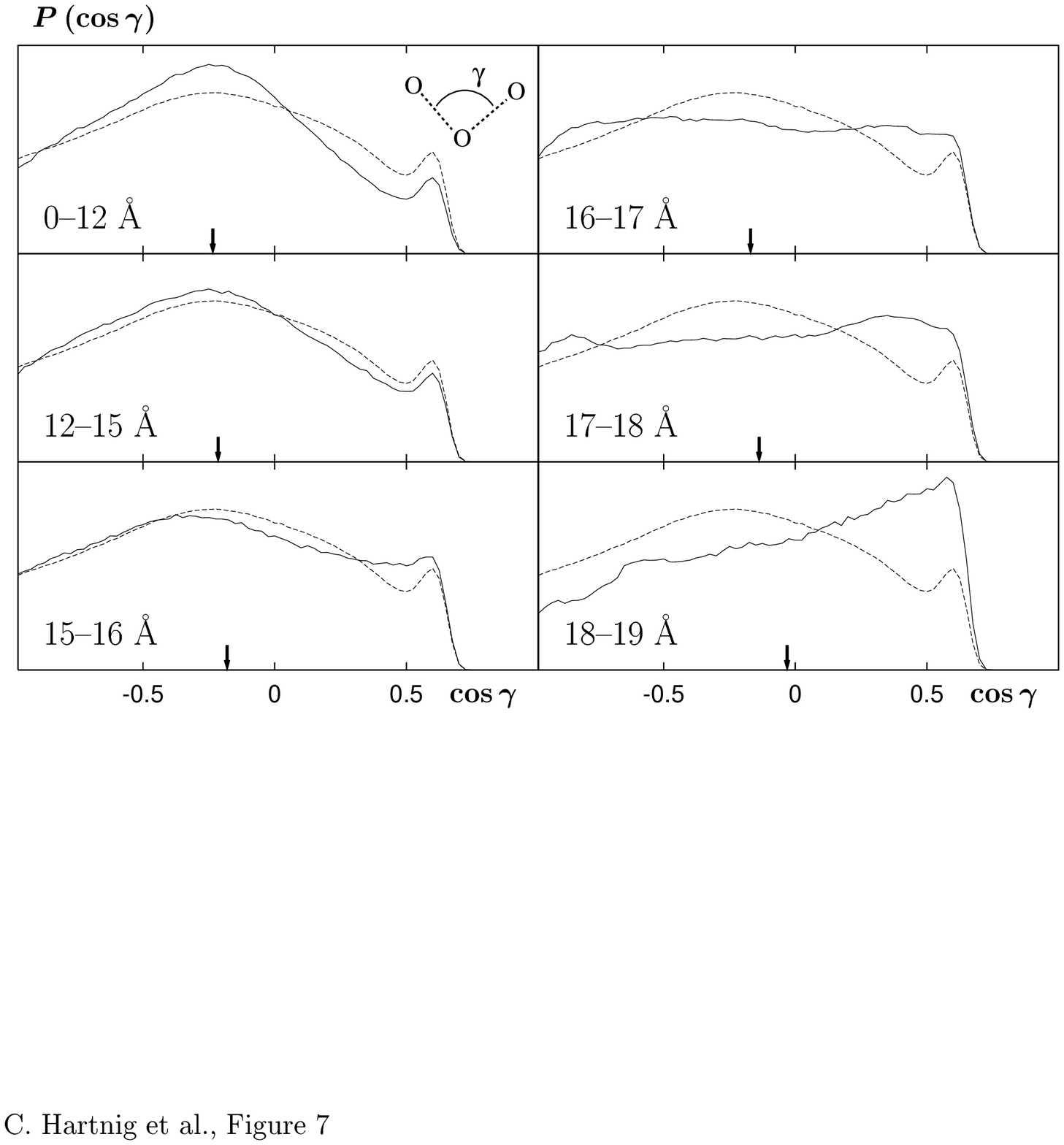,bbllx=38,bblly=358,bburx=560,bbury=713,clip=,scale=0.8}
    \caption{Distribution function of the oxygen-oxygen-oxygen angle,
      $P(\cos \gamma)$, (see inset)
      for different cylindrical shells as indicated. The dashed line is the 
      corresponding distribution in bulk water. Arrays indicate the average
      value, $\langle \cos\gamma\rangle$.}
    \label{fig:gamma}
    \figadjust
    \figadjust
  \end{center}
\end{figure}
We have further analyzed the local structure around a water molecule.
We explored the arrangement of nearest neighbor molecules by means of
the distribution of the angle $\gamma$, which is the angle between the
two vectors from an oxygen atom of a water molecule with the oxygen
atoms of the two closest water neighbors (see inset of
Fig.~\ref{fig:gamma}).  Figure~\ref{fig:gamma} shows the angular
distribution function $P(\cos \gamma)$ in the fully hydrated system
with $N_W=2600$ for different radial layers (full lines, as
indicated).  For comparison, the corresponding distribution for bulk
water is given as the dashed lines.  The arrows point to the average
value of $\cos\gamma$.  Both in experiment and in simulation, the
distribution function for bulk water at ambient conditions shows a
well defined peak at $\gamma \approx 105^\circ $ ($\cos\gamma \approx
-0.26$), representing the tetrahedral order, and a secondary peak at
$\gamma \approx 54^o$ ($\cos\gamma \approx 0.6$), attributed to
neighbors located in the cavities of the hydrogen bond
network~\cite{jedlovsky}.

In Vycor glass the distributions $P(\cos\gamma)$ near the center of
the pore are similar to the bulk distribution. In the distribution in
the innermost region with density less than ambient density, a more
ordered tetrahedral structure than in the bulk phase is quite obvious.
Closer to the boundary of the pore, the minimum between the main peak
of the tetrahedral coordination and the peak at $\cos\gamma = 0.6$
begins to fill in and for $\rho > 16$ {\AA} the distribution is
monotonic, indicating a departure from the preferential tetrahedral
arrangement.  In the layer between 18 and 19 {\AA}, there is an
obvious preference for small angles.  The average value, $\langle
\cos\gamma\rangle$ (arrows in Fig.~\ref{fig:gamma}), shifts to larger
values from the center to the surface of the pore.  The shift to
larger values is a consequence both of the substrate-induced
distortion of individual water-water hydrogen bonds and of the higher
local density, which leads to a collapse of the tetrahedral structure
due to packing effects.

\ekssection{Summary and Conclusions}

We have performed molecular dynamics simulations of water in porous Vycor glass
for different degrees of hydration, ranging from 19 to 96 \%. In the present work,
we analyzed the modifications of the hydrogen bond network of water at the
highest degree of hydration.  At nearly full
hydration the structural properties of water in the center of the pore are 
similar to the bulk phase. We did not compare the chemical potential of 
water in the pore to that of bulk water.

Close to the surface the structure changes in several ways.  Due to
the hydrophilic properties of the surface the density distribution of
hydrogen and oxygen atoms shows the formation of two layers of water
molecules which interact with the Vycor surface; the water-water
hydrogen bond network in the boundary region is distorted, as hydrogen
bonds between water molecules are partially substituted by bonds from
water to bridging oxygens of the Vycor glass. Hydrogen bonds close to
the surface are still preferentially linear but show a larger
deviation from linearity than hydrogen bonds in the center of the pore
or in bulk water. The local tetrahedral arrangement of water
molecules, however, is destroyed close to the surface due to the
geometric confinement and the increased local density.

The presence of the pore surface also changes the radial pair
distribution functions relative to those of bulk water.  The minimum
of the oxygen-oxygen radial distribution function fills in with
respect to the bulk; this observation is familiar from water under
pressure, where it also indicates a decrease in hydrogen bond order.
The comparison of the pair correlation functions with experiment is
encouraging, and we believe that further studies at different
temperatures would be helpful for a deeper understanding of the
phenomena concerned with the modification of hydrogen bonding in
confined water.

\section*{Acknowledgments}
We gratefully acknowledge financial support by the Fonds der Chemischen Industrie
and  the Clothilde Eberhardt Foundation.

\newpage

\end{document}